\documentclass[notitlepage,12pt]{article}
\pdfoutput=1
\usepackage{hyperref}

\usepackage{color}
\usepackage{amssymb,amsmath,amsfonts}
\usepackage{epsfig}
\usepackage{enumitem}
\usepackage{graphicx}
\usepackage{cite}
\usepackage[affil-it]{authblk}

\def\clock{{\count0=\time
           \divide\count0 60
           \ifnum\count0<10 0\fi\the\count0
           \multiply\count0 -60 \advance\count0 \time
           :\ifnum\count0<10 0\fi \the\count0
         }}
\newcommand{\timestamp}{{\small\vbox{\hbox{\tt\jobname.tex}
\hbox{\the\day/\the\month/\the\year, \clock}}}}

\newcommand{\beq}{\begin{equation}}
\newcommand{\eeq}{\end{equation}}
\newcommand{\ben}{\begin{displaymath}}
\newcommand{\een}{\end{displaymath}}
\newcommand{\beqa}{\begin{eqnarray}}
\newcommand{\eeqa}{\end{eqnarray}}
\newcommand{\bea}{\begin{eqnarray}}
\newcommand{\eea}{\end{eqnarray}}
\newcommand{\bean}{\begin{eqnarray*}}
\newcommand{\eean}{\end{eqnarray*}}
\newcommand{\ba}{\begin{array}}
\newcommand{\ea}{\end{array}}
\newcommand{\bi}{\begin{itemize}}
\newcommand{\ei}{\end{itemize}}

\newcommand{\lp}{\left(}
\newcommand{\rp}{\right)}

\numberwithin{equation}{section}
\begin{document}
\thispagestyle{empty}
\begin{titlepage}

\title{\textbf{Two Wave Functions and dS/CFT on $S^1\times S^2$}}
\author{Gabriele Conti\footnote{gabriele.conti@fys.kuleuven.be} , Thomas Hertog\footnote{thomas.hertog@fys.kuleuven.be}}
\affil{Institute for Theoretical Physics, KU Leuven, \\ Celestijnenlaan 200D, 3001 Leuven, Belgium}
\maketitle
\thispagestyle{empty}
\begin{abstract}

We evaluate the tunneling and Hartle-Hawking wave functions on $S^1\times S^2$ boundaries in Einstein gravity with a positive cosmological constant. In the large overall volume limit the classical predictions of both wave functions include an ensemble of Schwarzschild-de Sitter black holes. We show that the Hartle-Hawking tree level measure on the classical ensemble converges in the small $S^1$ limit. A divergence in this regime can be identified in the tunneling state. However we trace this to the contribution of an unphysical branch of saddle points associated with negative mass black holes. Using a representation in which all saddle points have an interior Euclidean anti-de Sitter region we also derive a holographic form of both semiclassical wave functions on $S^1\times S^2$ boundaries.

\end{abstract}
\end{titlepage}

\newpage

\setcounter{page}{1}
\tableofcontents
\newpage

\section{Introduction}

In cosmology one is interested in computing the probability measure for different classical configurations of geometry and fields on a spacelike surface. This measure is given by the universe's quantum state. Predictions for our observations are obtained from this by further conditioning on our observational situation and its possible location in each history, and then summing over what is unobserved \cite{Hartle2007,Hartle2010}.

Current models of the wave function of the universe such as Vilenkin's tunneling state \cite{Vilenkin1986,Vilenkin1987} or the Hartle-Hawking wave function \cite{Hartle1983} successfully predict that our classical universe emerged in an early period of inflation. In their present form however they are based on a weighting of four-geometries that is difficult to define beyond the semiclassical leading order in $\hbar$ approximation. It is an important goal of quantum gravity to obtain a precise formulation of the quantum state that can be used to reliably calculate the probability measure beyond the saddle point approximation. 

The dS/CFT correspondence \cite{Witten2001,Strominger2001,Balasubramanian2001,Maldacena2002} can be viewed as a program in this direction. Inspired by the Euclidean AdS/CFT duality \cite{Horowitz2003}, some versions of dS/CFT postulate that the asymptotic wave function of the universe is given in terms of the partition function of deformations of CFTs on ${\cal I}^{+}$ \cite{Maldacena2002,Garriga2008,Hertog2011,Anninos2012,Castro2012,Banerjee2013}. This idea has been realised concretely e.g. in the semiclassical Hartle-Hawking state for general configurations on $S^3$, where the holographic form of the wave function involves the partition function of certain complex deformations of Euclidean CFTs familiar from AdS/CFT \cite{Hertog2011}. The connection with asymptotic Euclidean AdS spaces comes about because the Hartle-Hawking saddle points in gravity coupled to a positive cosmological constant and a positive scalar potential have a representation in which the geometry consists of a regular Euclidean AdS domain wall that makes a smooth (but complex) transition to a Lorentzian, inflationary universe that is asymptotically de Sitter \cite{Maldacena2002,Hertog2011,Harlow2011,Hartle2012,Hartle2012b,Castro2012} (see also \cite{McFadden2009}). In this representation the tree level measure on the classical ensemble of histories is given by the regularised AdS domain wall action, which by AdS/CFT can be replaced by the logarithm of the partition function of a dual field theory\footnote{See e.g. \cite{Freivogel2006,Banks2011} for a different approach to holographic cosmology.}. Since the argument of the asymptotic wave function enters as an external source in the dual partition function this yields a holographic expression of the Hartle-Hawking semiclassical probability measure on the ensemble of asymptotically de Sitter histories.

Recently Anninos et al. \cite{Anninos2011} have put forward a precise realisation of dS/CFT that is potentially valid beyond the semiclassical approximation. Their proposal relates Vasiliev's theory of higher spin gravity in four-dimensional de Sitter space to a Euclidean, three dimensional conformal field theory with anti-commuting scalars and $Sp(N)$ symmetry. This has made possible the first precise holographic calculations of the wave function of the universe, by evaluating the partition function of the $Sp(N)$ CFT as a function of various deformations. It was found however that the resulting measure exhibits several divergences that are unexpected in well-defined, stable theories \cite{Anninos2012,Banerjee2013}. This includes divergences associated with mass deformations in the dual on $S^3$ and with the topological complexity on more complicated future boundaries. 

Evidently it is important to understand what this means and whether these divergences also occur in other theories in asymptotic de Sitter space\footnote{At least some of the divergences discussed in the context of Vasiliev gravity appear to be present also in Einstein gravity\cite{Anninos2012,Castro2012,Banerjee2013}.}. If so this potentially undermines the very notion of a wave function of the universe in quantum gravity. 
To elucidate these questions we perform a careful analysis of both the tunneling and the Hartle-Hawking wave function on $S^1\times S^2$ boundaries in Einstein gravity\footnote{See e.g. \cite{Bousso1998} for early work on the Hartle-Hawking wave function on $S^1 \times S^2$.}. In \cite{Anninos2012} a similar divergence was found in the small $S^1$ limit both in a bulk calculation in Einstein gravity and in a boundary calculation in the dual to Vasiliev gravity. Here we identify the former divergence in the tunneling state, but we find that the Hartle-Hawking measure converges at small $S^1$.

However we then analyse in detail the classical predictions of both wave functions and show that the divergence in the tunneling state is connected with the contribution of an unphysical branch of saddle points associated with negative mass black holes in de Sitter space. There are strong arguments that configurations describing negative mass black holes must be excluded from the physical configuration space in quantum gravity in order for the theory to be well-defined and stable. We show that discarding the corresponding saddle points branches renders both the tunneling and the Hartle-Hawking wave function in Einstein gravity on $S^1\times S^2$ well-behaved. Whether this is the correct approach in Vasiliev gravity, which may or may not be stable, remains to be seen.

The outline of this paper is as follows: In Section \ref{sect} we compute the tunneling wave function on $S^1\times S^2$ in the large overall volume limit, as a function of the relative size of $S^1$ and $S^2$. We rediscover the divergence in the small $S^1$ limit discussed in \cite{Anninos2012,Banerjee2013}. In Section \ref{sectnbwf} we compute the Hartle-Hawking wave function on $S^1\times S^2$ and find it converges at small $S^1$. In Section \ref{sectbc} we evaluate the wave functions at finite volume and, in particular, in the classically forbidden region. We show that classical evolution emerges only at exceedingly large overall volumes in the small $S^1$ limit. In Section \ref{sectcl} we derive the classical predictions of the asymptotic wave functions on $S^1 \times S^2$. We demonstrate that the divergence in the tunneling wave function is associated with a branch of saddle points describing negative mass Schwarzschild-de Sitter black holes.  In Section \ref{hwfs} we derive a holographic formulation of the semiclassical Hartle-Hawking wave function on $S^1\times S^2$ and clarify its connection with a Euclidean AdS wave function. We close with a discussion in Section \ref{disc}.

\section{Asymptotic Tunneling Wave Function} \label{sect}

In quantum cosmology in the semiclassical approximation the state of the universe is given by a wave function $\Psi$ defined on the superspace of all possible three-geometries and matter field configurations. All wave functions must satisfy the operator implementation of the classical Hamiltonian constraint known as the Wheeler-DeWitt (WDW) equation ${\cal H}\Psi=0$, where ${\cal H}$ is a differential operator on superspace. To solve the WDW equation one has to specify a set of boundary conditions on $\Psi$. This is the analogue in quantum cosmology of specifying the initial conditions for the universe.

Vilenkin has proposed \cite{Vilenkin1986,Vilenkin1987} that $\Psi$ consists of outgoing waves only at singular boundaries of superspace. The physical idea behind this proposal is that the universe originates in a non-singular quantum tunneling event. Vilenkin's tunneling proposal can be implemented as a boundary condition on the WDW equation. In the semiclassical approximation in which the wave function is written as a sum of terms of the form
\beq \label{tundefi}
\Psi_T = \sum_n A_n e^{iS_L^n},
\eeq
the tunneling boundary condition amounts to a positivity condition on the conserved current $J_n = \frac{i}{2} \Psi^*_T \stackrel{\leftrightarrow}{\nabla} \Psi_T = -|A_n|^2 \nabla S^n_L$ associated with WDW evolution. 

In a cosmological context, for spherical boundaries and in a minisuperspace approximation, the semiclassical tunneling wave function predicts an ensemble of classical, expanding universes with an early period of inflation \cite{Vilenkin1986,Vilenkin1987}. 
Here we are interested in $\Psi_T$ on $S^2 \times S^1$ boundaries in four dimensional Einstein gravity coupled to a positive cosmological constant $\Lambda$. The Lorentzian action of this model is given by\footnote{We use Planck units where $\hbar=G=c=1$.}
\beq \label{lact1}
S_L=\frac{1}{16 \pi} \int_{\mathcal{M}} d^4x \sqrt{-g}\lp R-2\Lambda\rp + \frac{1}{8\pi}\int_{\partial \mathcal{M}} d^3x \sqrt{h} K \,.
\eeq
To find $\Psi_T$ we evaluate the action on regular solutions with a boundary geometry of the form\footnote{Our calculations in this Section closely follow \cite{Anninos2012,Banerjee2013}.}
\beq \label{boundl}
R_c^2 \gamma_{ij}dx^idx^j=R^2_c\lp \lp \frac{\beta d\theta}{2\pi}\rp ^2+d\Omega_2^2\rp\,.
\eeq  
Hence the minisuperspace of this model is the two-dimensional manifold $0\leq R_c \leq\infty, \ 0\leq \beta \leq \infty$. The four-dimensional `saddle point' solutions that match onto boundaries of this form are complex generalizations of Schwarzschild -- de Sitter space. They can be written as
\beq \label{lks}
ds^2= -\rho ^2(\chi) d\lambda^2 +d\chi^2 + R^2(\chi)d\Omega _2 ^2 \,,
\eeq 
where $\chi$ is a complex variable that runs from $\chi_0$ at the `South Pole' (SP) of the saddle points, where the geometry closes off, to $\chi_c$ at the boundary where $R(\chi_c) = R_c$ and $\rho(\chi_c) = \rho_c$. To match the periodicity of the $S^1$ at the boundary \eqref{boundl} the variable $\lambda$ must be periodic with a period $\lambda_0$ given by
\beq \label{lambl}
\lambda_0 = \frac{\beta R_c}{\sqrt{-\rho^2_c}}.
\eeq
The equations of motion for the scale factors $\rho(\chi)$ and $R(\chi)$ are
\beqa \label{eoml}
\frac{\ddot{\rho}}{\rho}R+\ddot{R}+\frac{\dot{\rho}}{\rho}\dot{R}+\Lambda R=0 \nonumber \\
2R\ddot{R}+\dot{R}^2-1+\Lambda R^2=0\nonumber \\
\dot{R}^2+2\frac{\dot{\rho}}{\rho}\dot{R}R-1+\Lambda R^2=0
\eeqa
where $\dot R \equiv \partial_{\chi} R$.

We concentrate on solutions where the $S^1$ shrinks to zero size at the South Pole, at a nonzero value of the $S^2$ radius $R(\chi_0) = R_0$. This class of solutions determines the $\beta$-dependence of $\Psi_T$, which is our main focus in this paper\footnote{When the $S^2$ shrinks to zero size regularity at the SP implies that the saddle point is a quotient of de Sitter space with an amplitude that is independent of $\beta$. Hence this class of solutions merely accounts for an overall normalisation factor.}.
Regularity of the solutions at the SP implies
\beq \label{lambcil}
\lambda_0=\pm \frac{2\pi i}{\dot{\rho}(\chi _0)} = \pm \frac{4 \pi i R_0}{1- \Lambda R_0^2}\ .
\eeq
The second equality follows from the equations of motion \eqref{eoml} which admit a first integral of the form
\beq\label{int}
\rho^2=\dot{R}^2=1-\frac{\alpha}{R}-\frac{\Lambda}{3} R^2\,,
\eeq
where the constant of integration $\alpha=R_0-(\Lambda/3) R_0^3$. Evaluating the action \eqref{lact1} on these solutions yields
\beq \label{act1l}
S_L =-\frac{\lambda_0}{12} \lp 9 R_0 - \Lambda R_0^3 - 12 R_c +4 \Lambda R_c^3 \rp \ .
\eeq 

\begin{figure}[t!] 
\begin{center}
\includegraphics[scale=0.63]{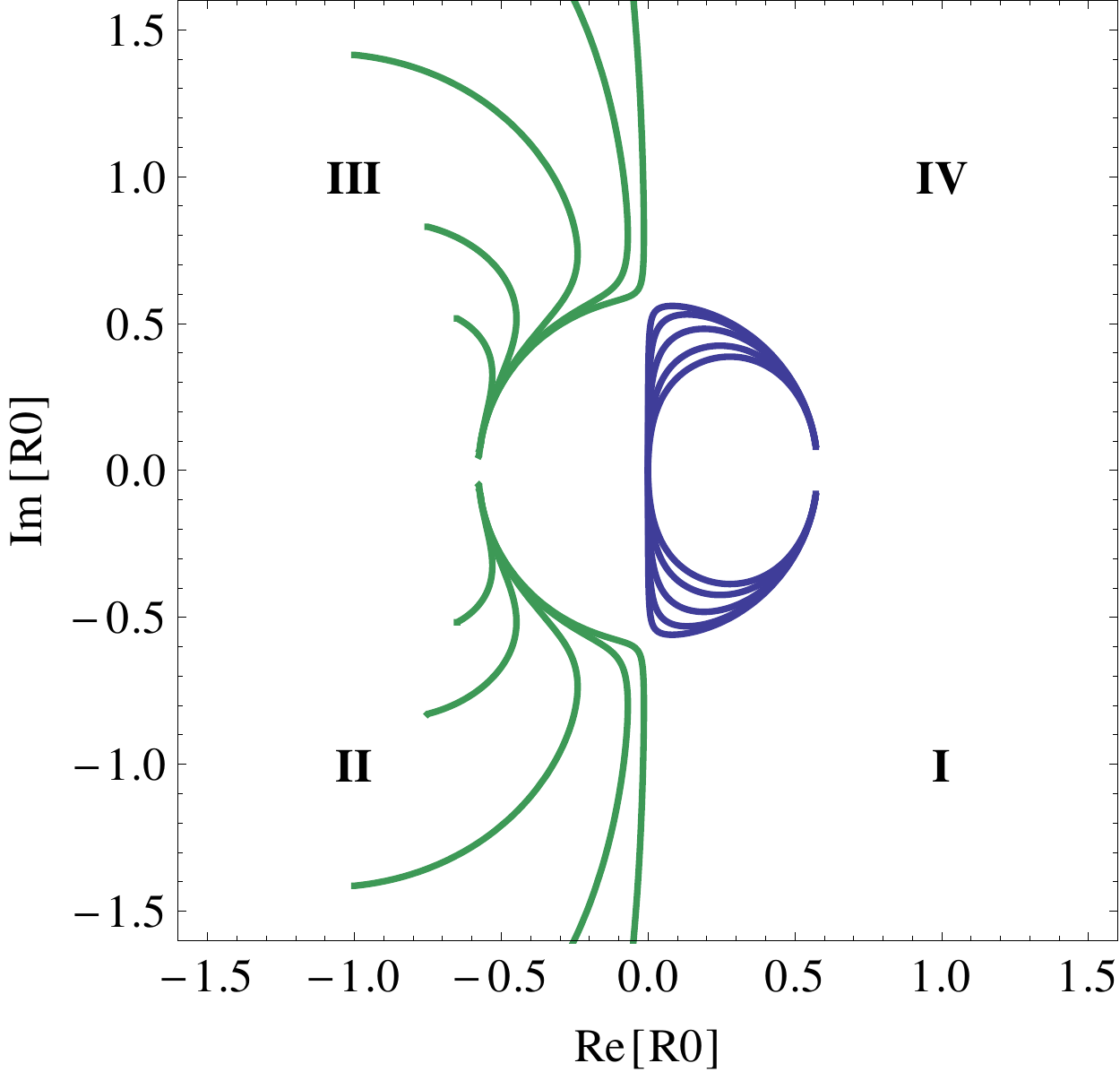}
\end{center}
\caption{The saddle points can be labeled by the complex value $R_0$ of the overall scale factor $R$ at the South Pole. They are naturally divided in four classes corresponding to values $R_0$ in different quadrants of the complex $R_0$-plane. For each boundary configuration $(R_c,\beta)$ in the classical region of superspace there is one saddle point in each quadrant that potentially contributes to the wave function. We show the behavior of $R_0$ as a function of $\beta$ (along the curves in each quadrant), for a number of values of the overall boundary radius $R_c$ in the classical region of superspace. As $\beta \rightarrow \infty$ we have $R_0 \rightarrow \pm 1/\sqrt{\Lambda}$, and $\Lambda=3$ here. In the $R_c \rightarrow \infty$ limit the four curves lie on the imaginary axis for $\beta\leq 2\pi / \sqrt{3}$ and on the circle at radius $|R_0|=1/\sqrt{\Lambda}$ for larger $\beta$.} 
\label{circle}
\end{figure}

The boundary value $R_c$ is real and positive of course, but $R_0$ is in general complex. The latter can be used to label the different saddle points. Combining \eqref{lambcil} with \eqref{lambl} and using \eqref{int} yields the following quintic equation for $R_0$ in terms of the argument $(\beta,R_c)$,
 \beq \label{quint}
 -16\pi^2 R_0^2 \lp -\frac{1}{R_c^2} + \frac{R_0-\frac{\Lambda}{3}R_0^3}{R_c^3}+\frac{\Lambda}{3}\rp = \beta ^2 (1-\Lambda R_0^2)^2\,.
\eeq
We discuss the solution of this equation for general values of the argument $(\beta,R_c)$ in Section \ref{sectbc}. Here we concentrate on the classical region of superspace $R_c \gg 1/\sqrt{\Lambda}$. In this region there are four classes of solutions corresponding to values $R_0$ in different quadrants of the complex $R_0$-plane. For each boundary configuration $(R_c,\beta)$ there is one saddle point in each quadrant. Figure \ref{circle} shows the behavior of $R_0$ as a function of $\beta$ in each quadrant, for a number of values of the overall boundary radius $R_c$ in the classical region of superspace.

In the $R_c \rightarrow \infty$ limit the four curves tend to curves that run along the imaginary axis for $\beta \leq2\pi / \sqrt{3}$ and along a circle of radius $|R_0|=1/\sqrt{\Lambda}$ for larger values of $\beta$. In this limit \eqref{quint} reduces to a quartic equation which can be solved analytically \cite{Anninos2012,Banerjee2013}, yielding
\beq \label{saddlesl}
R_0= -\frac{2 \pi}{\beta \Lambda} \lp  \rho_c/R_c \pm \sqrt{(\rho_c/R_c)^2 +\frac{\beta ^2}{4\pi ^2 \Lambda}}  \rp + \mathcal{O}\lp\frac{1}{R_c} \rp\ .
\eeq
The equations of motion together with the boundary conditions \eqref{lambcil} at the SP imply that $\rho_c/R_c \rightarrow -i \sqrt{\Lambda/3}$ in the large $R_c$ regime if $Im[R_0]>0$ and $\rho_c/R_c \rightarrow +i \sqrt{\Lambda/3}$ for saddles with $Im[R_0]< 0$.
Hence \eqref{saddlesl} describes the limiting (large $R_c$) behaviour of the four classes of solutions shown in Fig. \ref{circle}.
In the large $R_c$ regime eq. \eqref{lambl} becomes 
\beq \label{rat21}
\lambda_0 = \frac{\beta}{\sqrt{\Lambda/3}}\lp 1+\frac{3}{2\Lambda R_c^2} -\frac{3\alpha}{2 \Lambda R_c^3} \rp +\mathcal{O} \lp \frac{1}{R_c^4}\rp\ .
\eeq
Substituting this in the action \eqref{act1l} yields
\beq \label{actl}
iS_L = -i \frac{\beta}{\sqrt{\Lambda/3}} \lp \frac{\Lambda}{3} R_c^3 - \frac{1}{2}R_c +\frac{1}{4}R_0(1+ \frac{\Lambda}{3} R_0^2)\rp + \mathcal{O} \lp \frac{1}{R_c}\rp\,,
\eeq
with $R_0$ given by \eqref{saddlesl}. Hence the action obeys the positivity condition on the conserved current in the Lorentzian regime for all four classes of saddle points. It would therefore seem they all contribute to the tunneling wave function $\Psi_T$. 

At large overall volume the semiclassical wave function can be elegantly expressed in terms of a universal phase factor - which accounts for the `counterterms' in holographic discussions - multiplied by a sum of asymptotically finite `regularized' saddle point actions. From \eqref{actl} we obtain
\beq \label{Tresult}
\Psi_T [\beta,R_c]\propto  \frac{1}{2} e^{iS_{ct}} \left(\cosh(I^1_{R})e^{iS_R} + \cosh(I^2_{R}) e^{-iS_R} \right)
\eeq 
where 
\beq \label{count}
iS_{ct}(\beta,R_c)= -i \frac{\beta}{\sqrt{\Lambda/3}} \lp \frac{\Lambda}{3} R_c^3 - \frac{1}{2}R_c\rp \,,
\eeq
and
\beq
iS_R (\beta)= -i\lp \frac{\lp\beta ^2 -4\pi ^2\rp ^{3/2}}{9\Lambda \beta ^2}\rp
\label{phase}
\eeq
if $\beta > 2\pi/\sqrt{3}$, and $iS_R=0$ otherwise.
The amplitudes in eq. \eqref{Tresult} are given by
\bea \label{actionsl}
I^1_{R}(\beta) = - I^4_{R}(\beta) &=&\frac{4\pi^2}{9 \beta ^2 \Lambda} \mathrm{Re}\left[\lp -1 + \sqrt{1-\frac{3\beta ^2}{4 \pi ^2}}\rp \lp \frac{3\beta ^2}{2\pi}-\pi +\pi \sqrt{1-\frac{3\beta ^2}{4\pi^2}} \rp \right] \nonumber \\
I^2_{R}(\beta)  = - I^3_{R}(\beta) &=&\frac{4\pi^2}{9 \beta ^2 \Lambda} \mathrm{Re}\left[\lp 1 + \sqrt{1-\frac{3\beta ^2}{4 \pi ^2}}\rp \lp -\frac{3\beta ^2}{2\pi}+\pi +\pi \sqrt{1-\frac{3\beta ^2}{4\pi^2}} \rp \right]
\eea
where the superscript on $I_R$ in \eqref{actionsl} corresponds to the label of the quadrant in Fig \ref{circle}.   

For $\beta > 2\pi / \sqrt{3}$ we have $I^1_R = \bar I^2_R$ with imaginary part given by \eqref{phase} and real part given by 
\beq \label{lactsad}
I^1_R(\beta)  = -\lp \frac{\pi}{\Lambda} - \frac{8\pi ^3}{9\Lambda \beta ^2}\rp,
\eeq
which tends to the well known Nariai amplitude $-\pi/\Lambda$ as $\beta \rightarrow \infty$. Even though the phase of the wave function is dominated by the universal $iS_{ct}$ factor at large overall volume, we will see in Section \ref{sectcl} that the imaginary part of the regularized actions \eqref{lactsad} plays a crucial role on the classical predictions of the wave function.

For $\beta \leqslant 2\pi / \sqrt{3}$ the solutions $R_0$ are purely imaginary. The regularized actions \eqref{actionsl} are all real in this regime, but there is no obvious relation between $I^1_R$ and $I^2_R$. Indeed in the $\beta \rightarrow 0$ limit, 
\beq \label{divergence}
I^1_R \rightarrow 0\,, \hspace{1cm} I^2_R = \frac{16 \pi ^3}{9 \Lambda \beta ^2} \rightarrow  \infty\,.
\eeq
The second class of solutions gives rise to a diverging amplitude as observed in \cite{Anninos2012, Banerjee2013}. Here we have found that this divergence emerges as a feature of the tunneling wave function. We return to its interpretation below, but we first evaluate the Hartle--Hawking wave function on $S^1 \times S^2$.

\section{Asymptotic Hartle--Hawking Wave Function}\label{sectnbwf}

A different proposal for boundary conditions on the Wheeler-de Witt equation is due to Hartle and Hawking \cite{Hartle1983}
who have suggested, inspired by the Euclidean construction of the ground state wave function in field theory, that the wave function of the universe is given in terms of an appropriately defined Euclidean path integral. In the semiclassical approximation the Hartle--Hawking (HH) wave function is thus given by
\beq \label{nbwf}
\Psi_{HH}(\xi) \simeq \sum_n e^{-I^n_E[\xi]} 
\eeq
where $I^n_E$ is the Euclidean action of a compact, regular - and therefore generally complex - saddle point solution that matches the real boundary data $\xi$ on its only boundary. 
The sum over saddle points is such that the resulting wave function is real. 

The Euclidean action of the model we consider here reads
\beq \label{eact}
I_E=-\frac{1}{16 \pi} \int_{\mathcal{M}} d^4x \sqrt{g}\lp R-2\Lambda\rp - \frac{1}{8\pi}\int_{\partial \mathcal{M}} d^3x \sqrt{h} K \,.
\eeq
To evaluate the HH wave function on $S^1 \times S^2$ boundaries \eqref{boundl} we consider Euclidean four-geometries of the form
\beq \label{eks}
ds^2=d\chi ^2 +\rho ^2\lp \chi\rp d\omega ^2+R^2\lp \chi\rp d \Omega_2^2 \,.
\eeq
where, as before, the variable $\chi$ goes from $\chi_0$ at the SP where the $S^1$ smoothly caps off and $R(\chi_0) = R_0$, to $\chi_c$ at the boundary where $R(\chi_c) = R_c$. In order for the circle in \eqref{eks} to match the periodicity of the $S^1$ at the boundary \eqref{boundl} we must have
\beq \label{ombound}
\omega_0 = \frac{\beta R(\chi_c)}{\rho(\chi_c)}\,.
\eeq
The Euclidean action evaluated on solutions of the form \eqref{eks} is given by
\beq \label{euact}
I_E=\frac{\omega_0}{12} \lp 9 R_0 - \Lambda R_0^3 -12 R_c +4 \Lambda R_c^3 \rp \,.
\eeq
Smoothness at the SP requires $\omega$ to be periodic with periodicity
\beq \label{omega0}
\omega(\chi_0)\equiv \omega_0=\frac{2\pi}{\dot{\rho}(\chi _0)} = \frac{4 \pi R_0}{1-\Lambda R_0^2}\ ,
\eeq
where the last equality follows from the equations of motion. Combining this with \eqref{ombound} yields again the quintic equation \eqref{quint} for $R_0$ as a function of $R_c$ and $\beta$. Hence we obtain the same set of saddle points in the large $R_c$-limit as before, specified by the solutions \eqref{saddlesl}. Their relative weighting however will be different as we see below.

It follows from \eqref{omega0} that saddle points specified by complex conjugate values of $R_0$ have complex conjugate $\omega _0$. Using \eqref{ombound} this means that complex conjugate $R_0$ go together with complex conjugate values $\rho(\chi_c)$ (since $\beta$ and $R_c$ are real and positive). Expanding \eqref{ombound} for large $R_c$ thus gives 
\beq \label{rat2}
\omega_0 = \pm i \frac{\beta}{\sqrt{\Lambda/3}} \lp 1+\frac{3}{2\Lambda R_c^2}-\frac{3\alpha}{2 \Lambda R_c^3} \rp +\mathcal{O} \lp \frac{1}{R_c^4}\rp\,,
\eeq
where the overall plus sign corresponds to saddles with $Im[R_0]>0$ and vice versa.
Substituting this in the Euclidean action \eqref{euact} yields
\beq \label{act}
I_E = \pm i \frac{\beta}{\sqrt{\Lambda/3}} \lp \frac{\Lambda}{3} R_c^3 - \frac{1}{2}R_c +\frac{1}{4}R_0(1+ \frac{\Lambda}{3} R_0^2)\rp + \mathcal{O} \lp \frac{1}{R_c}\rp\,,
\eeq 
where the overall plus sign again corresponds to saddle points with $Im[R_0]>0$. Summing the contributions of all four saddle points yields the following result for the semiclassical HH wave function evaluated on $S^1 \times S^2$, 
\beq \label{HHresult}
\Psi_{HH} [\beta,R_c] \propto e^{-I^1_R} \cos [S_{ct}+S_R] + e^{-I^2_R}  \cos [S_{ct}-S_R] 
\eeq
where $S_{ct}$ and $S_R$ are given by \eqref{count} and \eqref{phase}, and  
$I^1_R$ and $I^2_R$ are given by \eqref{actionsl}. 

The semiclassical HH wave function \eqref{HHresult} is manifestly real as expected.
Its behaviour in the regime $\beta > 2\pi/\sqrt{3}$ follows directly from \eqref{lactsad}. The usual Nariai limit in which $I^1_R =I^2_R =-\pi/\Lambda$ emerges as $\beta\rightarrow \infty$. At the critical value $\beta \equiv \beta_c = 2\pi/\sqrt{3}$ the amplitude of both terms in \eqref{HHresult} is given by $I_R = -\pi/(3\Lambda)$. At low $\beta$ the HH wave function behaves very differently from the tunneling wave function \eqref{Tresult}. Whereas $\Psi_T$ diverges in this limit, the HH wave function is manifestly well-behaved since $I^1_R \rightarrow 0$ and $I^2_R \rightarrow \infty$ as $\beta \rightarrow 0$.

\section{Wave Functions in the Classically Forbidden Regime}\label{sectbc}

We now proceed to evaluate $\Psi_T$ and $\Psi_{HH}$ for all values of $R_c$ and in particular in the classically forbidden region of superspace $0 \leq R_c\leq 1/\sqrt{\Lambda}$ where the superpotential 
\beq
\label{super}
U\lp R_c, \beta \rp = R_c \beta (1-\Lambda R_c^2) 
\eeq
is positive. This is the regime where the wave functions don't oscillate - and hence cannot be interpreted in terms of (an ensemble of) classical histories - but where they either grow or decay. This is also where the difference between the boundary conditions on $\Psi_T$ and $\Psi_{HH}$ becomes most manifest.

The semiclassical tunneling wave function is specified by the boundary condition that the wave function consists of outgoing modes only in the classical region of superspace $R_c \gg1/\sqrt{\Lambda}$. Therefore to evaluate $\Psi_T$ at finite values of $R_c$ and in particular in the classically forbidden region we start from its large $R_c$ form \eqref{Tresult} and solve \eqref{quint} numerically to find the wave function at smaller values of $R_c$.

\begin{figure}[t!] 
\begin{center}
\includegraphics[scale=.45]{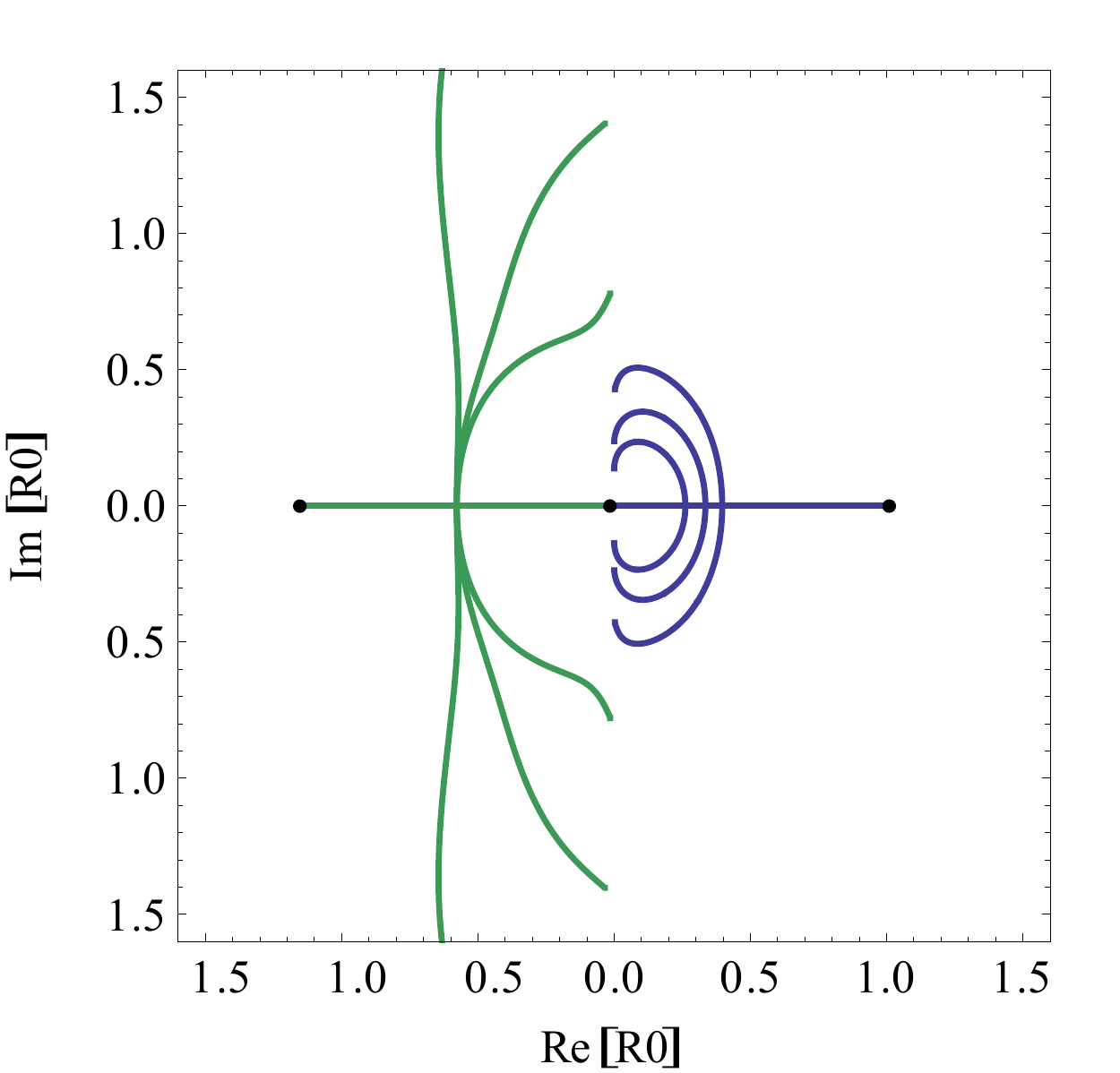} \hspace{1cm}
\includegraphics[scale=.45]{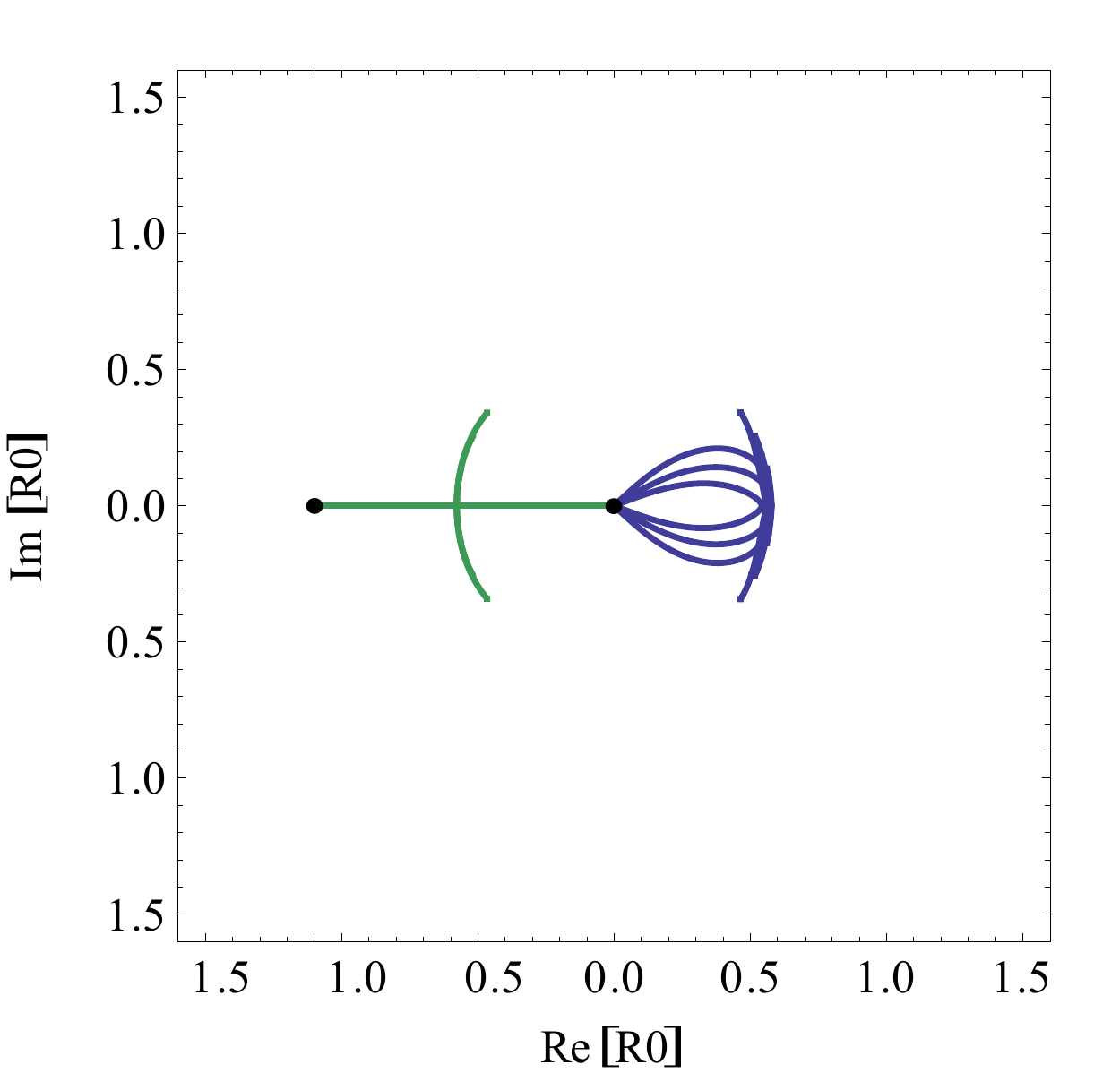} 
\end{center}
\caption{The trajectories in the complex $R_0$-plane that specify the four saddle points contributing to $\Psi_{T}$ as a function of the boundary value $R_c$, for a number of different values of $\beta$. The left panel shows the four solutions $R_0(R_c)$ for three values of $\beta$ in the high temperature regime $\beta\leq 2\pi/\sqrt{3}$ in which $R_0$ tends to a purely imaginary value as $R_c \rightarrow \infty$. The right panel shows $R_0(R_c)$ for three values of $\beta$ in the low temperature regime $\beta > 2\pi/\sqrt{3}$ in which the four curves tend to a point on the circle $|R_0|=1/\sqrt{\Lambda}$ in the $R_c \rightarrow \infty$ limit.}
\label{R0tun}
\end{figure}

For each value of $\beta$ this yields four curves $R_0(R_c)$ in the complex $R_0$-plane, corresponding to four families of saddle points. As $R_c$ increases the curves tend to two pairs of complex conjugate values $R_0$ that define $\Psi_T$ in the $R_c \rightarrow \infty$ limit as discussed above. Fig. \ref{R0tun} shows several examples of saddle point trajectories $R_0(R_c)$ in the complex $R_0$-plane, for a number of different values of $\beta$. 

As before we can clearly distinguish a low - and a high temperature regime.
If $\beta\leq \frac{2\pi}{\sqrt{3}}$ the trajectories start out somewhere on the imaginary axis in the $R_c \rightarrow \infty$ limit. At a critical boundary radius $R_c \sim1/\sqrt{\Lambda}$ each pair of complex conjugate solutions tends to a real solution\footnote{The precise value of $R_c$ at which the transition from complex to real saddle points occurs slightly depends on $\beta$ and on the branch.}. The latter then becomes a pair of real solutions as $R_c$ decreases further, with $R_0 \rightarrow \pm \sqrt{3/\Lambda}$ and $R_0 \rightarrow 0$ in the $R_c \rightarrow 0$ limit.

If $\beta > \frac{2\pi}{\sqrt{3}}$ the trajectories start out somewhere on the circle in Fig. \ref{circle}. Other than this their behaviour is rather similar to the low $\beta$ regime, except that the branch of solutions with $\mathrm{Re}[R_0]>0$ is specified by complex $R_0$
over the entire range of radii $R_c$.

\begin{figure}[t!] 
\begin{center}
\includegraphics[scale=.6]{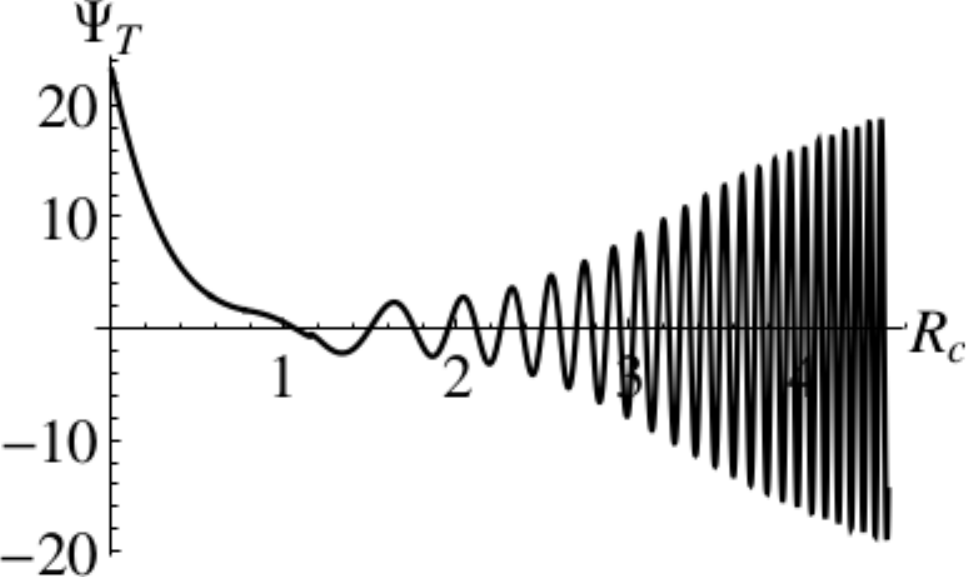}\hspace{1cm}
\includegraphics[scale=.6]{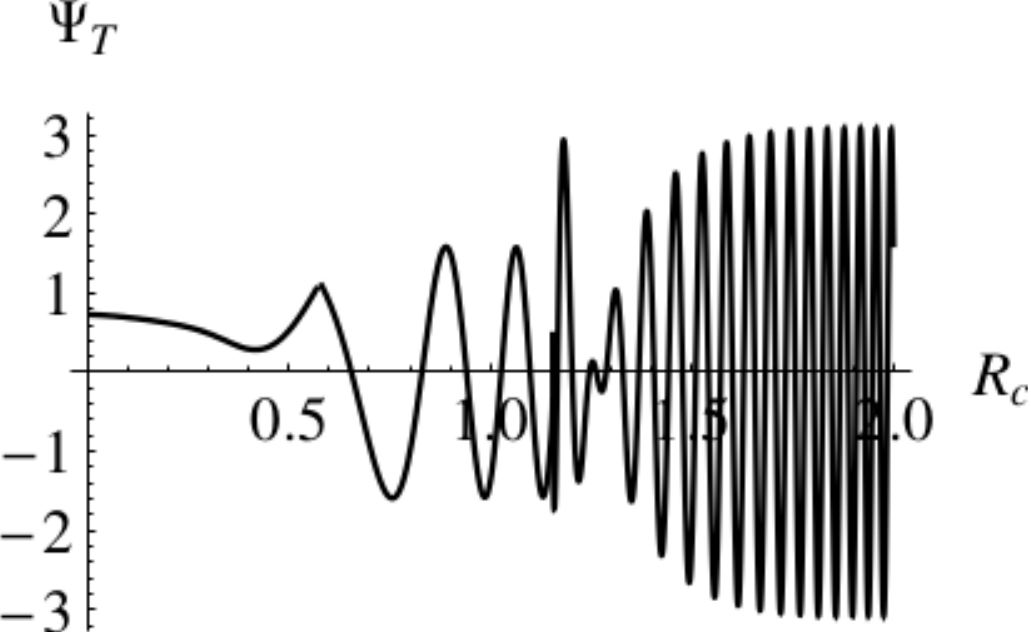}
\end{center}
\caption{The semiclassical tunneling wave function evaluated on $S^1 \times S^2$ as a function of the overall scale factor $R_c$, for two different (representative) values of the relative size $\beta$, i.e. $\beta=1.6$ (left) and $\beta = 15$ (right). As $\beta$ decreases the amplitude of the asymptotic classical configuration increases. }
\label{tunRc}
\end{figure}

To find $\Psi_{T}(R_c,\beta)$ we evaluate the action \eqref{act1l} on the above solutions $R_0(R_c,\beta)$ and sum over the different saddle points. As eq. \eqref{Tresult} shows in the large $R_c$ regime this yields a superposition of two outgoing waves. Each wave receives contributions from a set of saddle points with complex conjugate $R_0$. 

At the boundary of the classical region - or more precisely, at the critical value of the boundary radius $R_c$ where the saddle points become real - we use the WKB connection formulae to find the wave function under the barrier. The latter takes the form \cite{Vilenkin1987}
\beq \label{split}
\Psi_{T}(R_c,\beta) = \sum_{j=1,2} \left( \Psi^j_{+}(R_c,\beta) - \frac{i}{2} \Psi^j_{-}(R_c,\beta) \right).
\eeq
Here the index $j$ labels the individual terms (waves) in \eqref{Tresult}. That is, the linear combinations in \eqref{split} are matched, for each $j$, onto an outgoing wave in the classical region. The subscript $+/-$ refers to the leading/subleading saddle point under the barrier. The solutions $\Psi_{\pm}$ in the classically forbidden region are approximately real. 

We illustrate the resulting behaviour of the tunneling wave function in Fig.\ref{tunRc} for two representative values of $\beta$, namely $\beta=1.6$ as an example in the high temperature regime and $\beta = 15$ as an example at low temperature. One clearly sees that the oscillatory behaviour characteristic of the classical WKB regime only emerges at sufficiently large boundary values $R_c \gg 1/\sqrt{\Lambda}$ - well inside the classical region of superspace. The amplitude of the outgoing wave increases for decreasing $\beta$, and diverges for $\beta \rightarrow 0$ as we discussed above. In the classically forbidden region the wave function grows under the barrier in a way reminiscent of the tunneling wave function on $S^3$ boundaries in a cosmological context \cite{Vilenkin1987}.

We now turn to the Hartle--Hawking Wave Function in the classically forbidden region.
A defining feature of $\Psi_{HH}$ is the boundary condition that the wave function decays in the $R_c \rightarrow 0$ limit. In the saddle point approximation this selects the branches of saddle points for which $R_0 \rightarrow 0$ as $R_c \rightarrow 0$. In terms of the wave function, this selects only the $\Psi^j_{-}$ WKB components under the barrier.

Therefore to find the Hartle-Hawking wave function at finite nonzero values of $R_c$ we numerically solve \eqref{quint} and select the branches of solutions which yield a semiclassical wave function that obeys the above boundary condition as $R_c \rightarrow 0$.  As expected from the form of the superpotential \eqref{super} we find $\Psi_{HH}$ decays under the barrier and oscillates with an approximately constant amplitude given by \eqref{HHresult} in the region $R_c \gg 1/\sqrt{\Lambda}$.

\begin{figure}[t] 
\begin{center}
\includegraphics[scale=.45]{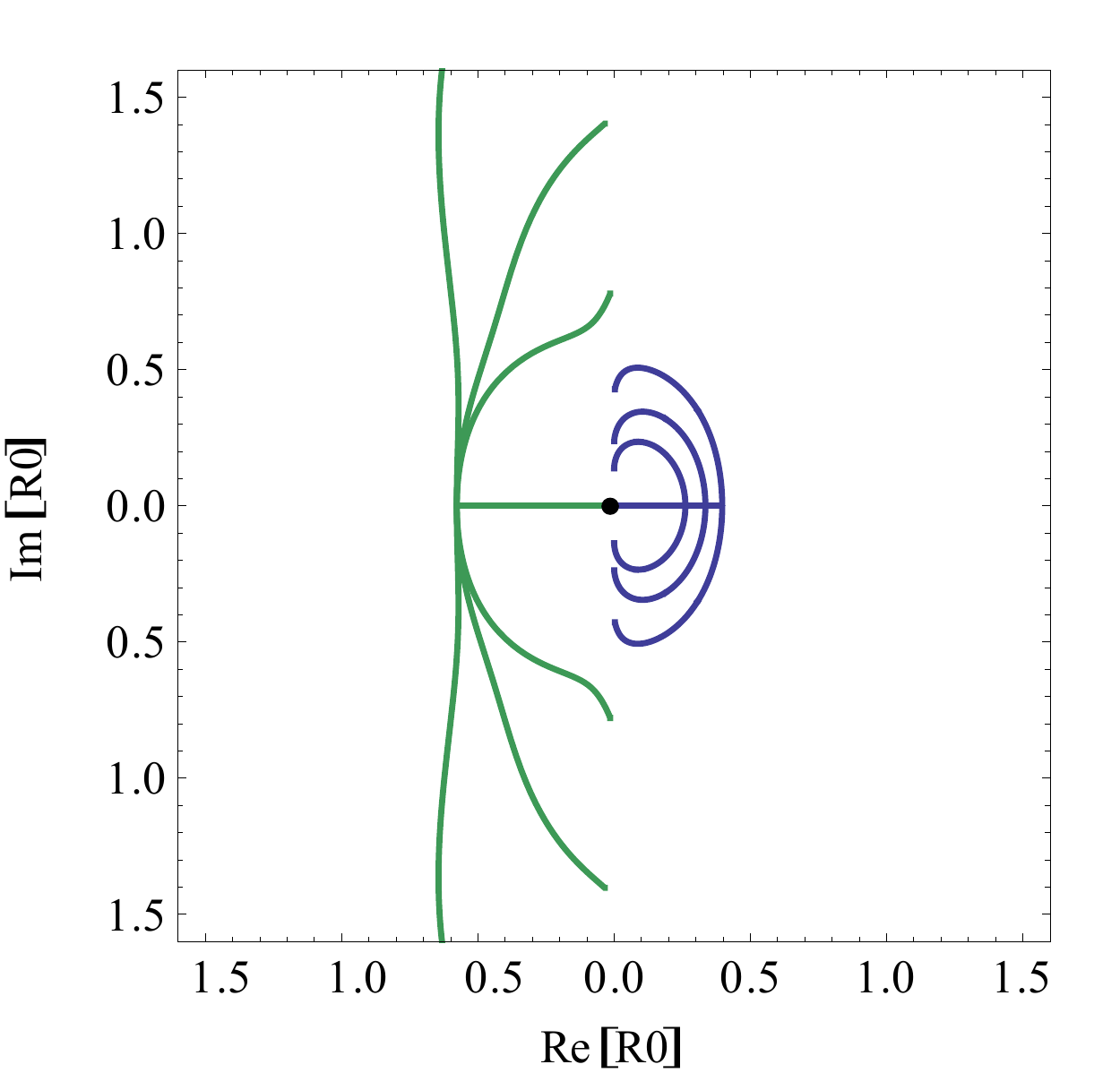}\hspace{1cm}
\includegraphics[scale=.45]{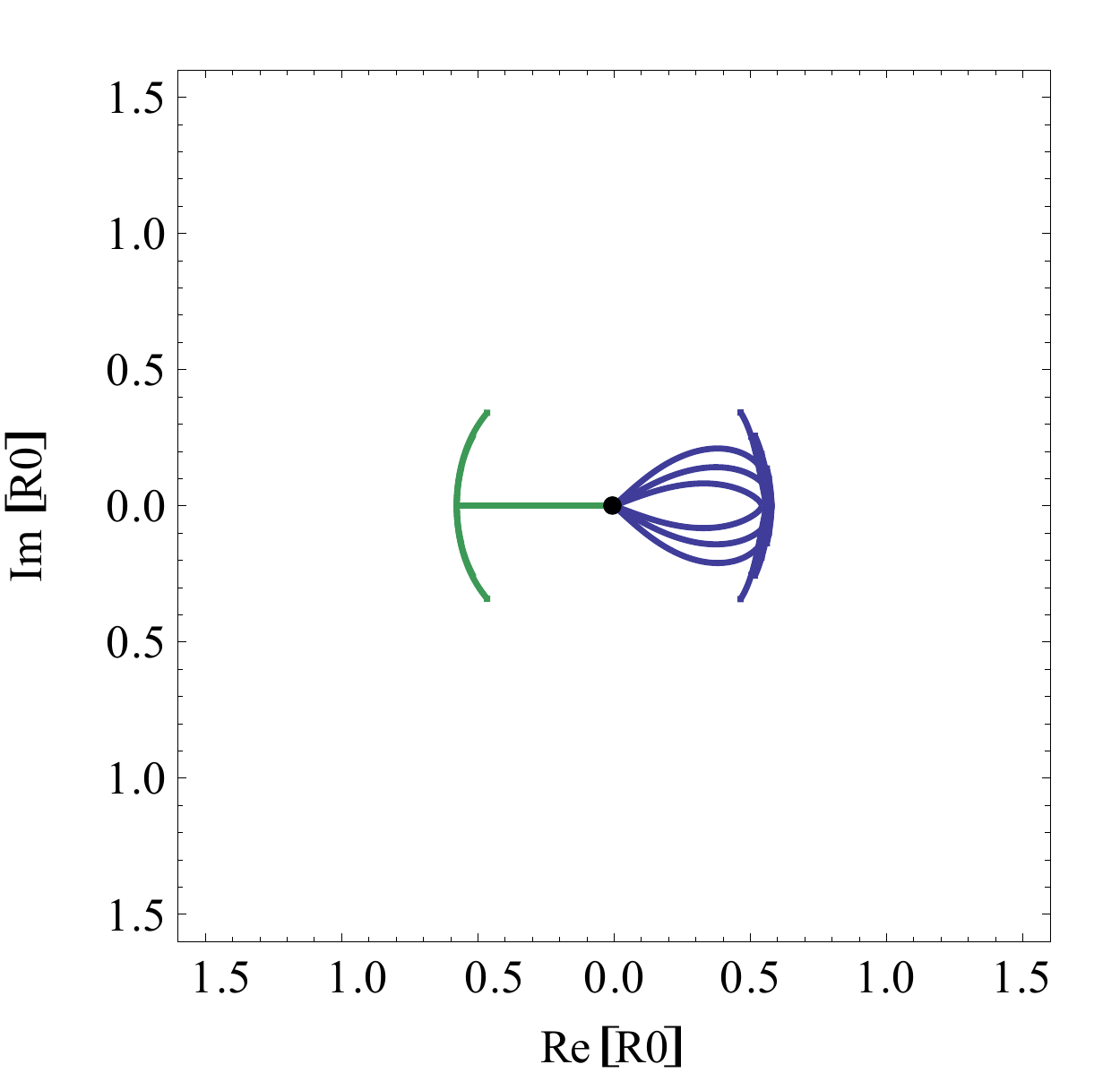}
\end{center}
\caption{The trajectories in the complex $R_0$-plane that specify the saddle points contributing to $\Psi_{HH}$ as a function of the boundary value $R_c$, for a number of different values of $\beta$. The left panel shows the solutions $R_0(R_c)$ for three values of $\beta$ in the regime $\beta\leq 2\pi/\sqrt{3}$ in which $R_0$ tends to a purely imaginary value as $R_c \rightarrow \infty$. The right panel shows $R_0(R_c)$ for three values of $\beta$ in the regime $\beta > 2\pi/\sqrt{3}$ in which the curves tend to a point on the circle $|R_0|=1/\sqrt{\Lambda}$ in the $R_c \rightarrow \infty$ limit.}
\label{R0beta}
\end{figure}

The wave function also depends on $\beta$. As before we can clearly distinguish a low - and a high temperature regime.
For $\beta\leq 2\pi/\sqrt{3}$ the Hartle-Hawking boundary condition selects two saddle point solutions in the small $R_c$ regime which are specified by real values of $R_0$. At a critical boundary radius $R_c$ around $1/\sqrt{\Lambda}$ each real solution splits\footnote{The exact value of $R_c$ at which this happens depends on $\beta$ and on the branch.} in a pair of saddle points specified by complex conjugate values of $R_0$. We use the WKB matching conditions at that point to obtain the wave function at larger values of $R_c$ where we recover the oscillating wave function \eqref{HHresult} in the region $R_c \gg 1/\sqrt{\Lambda}$. 

We show the saddle point trajectories $R_0 (R_c)$ in the complex $R_0$-plane in Fig. \ref{R0beta}(a), for four values of $\beta\leq 2\pi/\sqrt{3}$. The trajectories of course tend to two pairs of complex conjugate points on the imaginary axis as $R_c \rightarrow \infty$. In the $\beta \rightarrow 0$ limit the limiting points have $|R_0| \rightarrow \infty$ for one pair of solutions and $|R_0| \rightarrow 0$ for the second pair.

The behaviour of $R_0 (R_c)$ in the low temperature regime $\beta>\frac{2\pi}{\sqrt{3}}$ is summarised in Fig. \ref{R0beta}(b). The trajectories of the branch that starts out at $R_0 <0$ for small $R_c$ tend to a point on the $|R_0| = 1/\sqrt{\Lambda}$ circle away from the imaginary axis as $R_c \rightarrow \infty$. The second pair of solutions has complex $R_0$ for all values of $R_c >0$, except at $R_c=0$ and at $R_c=1/\sqrt{\Lambda}$ when $R_0=1/\sqrt{\Lambda}$.
 
\begin{figure}[t] 
\begin{center}
\includegraphics[scale=0.6]{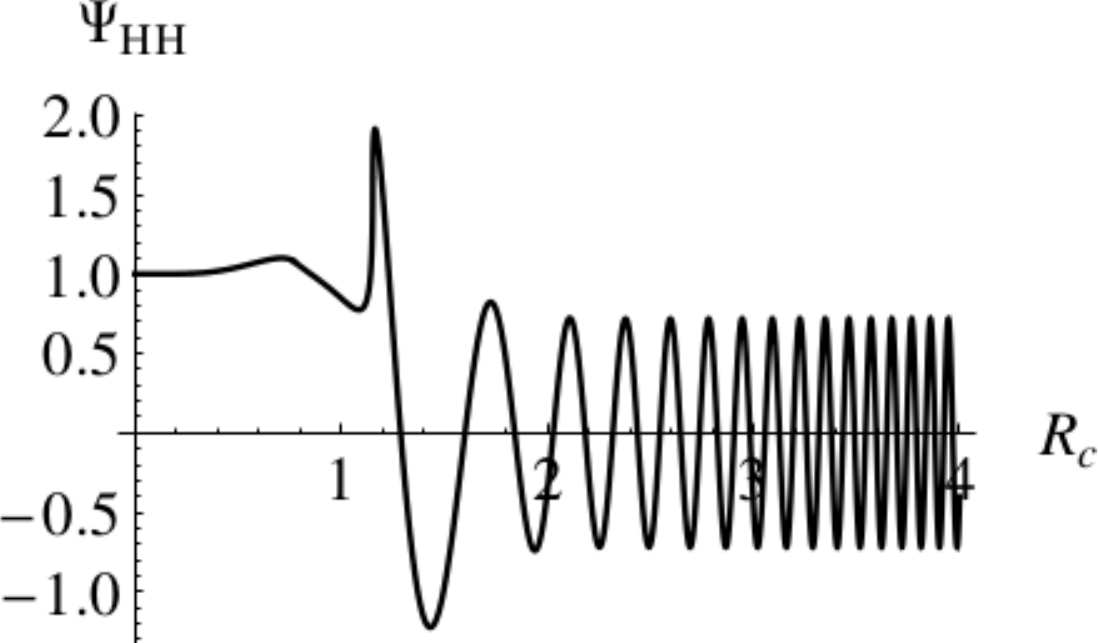}\hspace{1cm}
\includegraphics[scale=0.6]{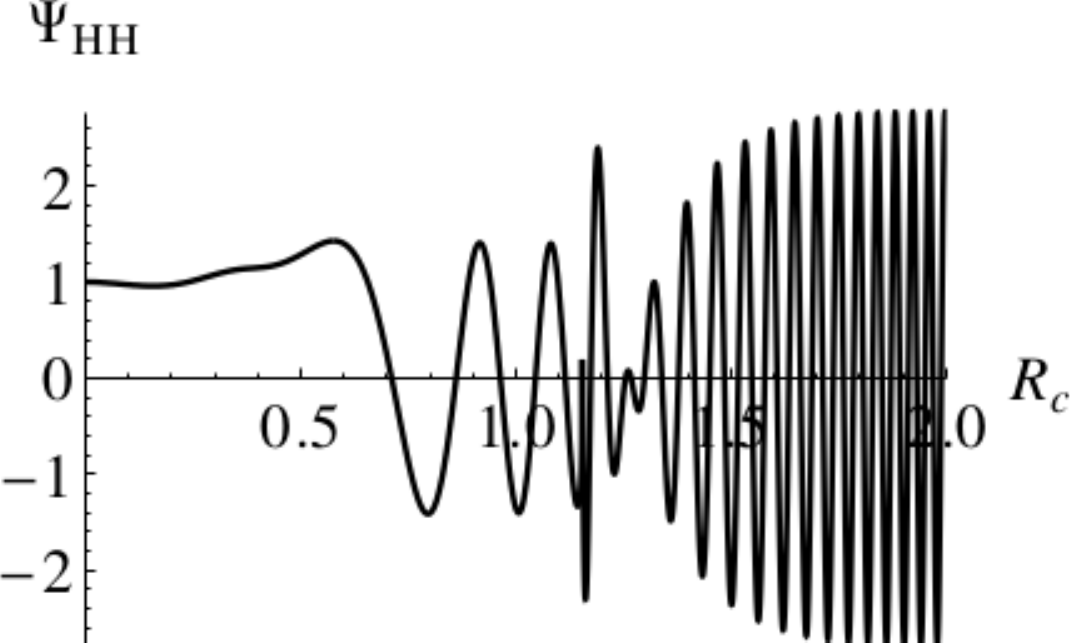}
\end{center}
\caption{The semiclassical Hartle--Hawking wave function evaluated on $S^1 \times S^2$ as a function of the overall scale factor $R_c$, for two different (representative) values of the `temperature' $\beta$, i.e.$\beta=1.6$ (left) and $\beta = 15$ (right). The wave function oscillates when $R_c \gg 1/\sqrt{\Lambda}$ where it predicts an ensemble of Lorentzian, classical Schwarzschild - de Sitter spaces with relative probabilities given by the amplitude of the wave.}
\label{nbwfbc}
\end{figure}

To find $\Psi_{HH}(R_c,\beta)$ we evaluate the Euclidean action \eqref{euact} on the above solutions $R_0(R_c,\beta)$ and sum over the different saddle points. The resulting wave function is shown in Fig. \ref{nbwfbc} for two representative values of $\beta$, namely $\beta=1.6$ as an example in the high temperature regime and $\beta = 15$ as an example at low temperature. One clearly sees that the oscillatory behaviour characteristic of the classical WKB regime only emerges at sufficiently large boundary values $R_c \gg 1/\sqrt{\Lambda}$ - well inside the classical region of superspace. The asymptotic amplitude of $\Psi_{HH}$ increases with increasing $\beta$ and tends to the Nariai amplitude $e^{\frac{\pi}{\Lambda}}$ in the $\beta\rightarrow \infty$ limit. For boundary values $R_c < 1/\sqrt{\Lambda}$ in the classically forbidden region the wave function is given by a sum of approximately real saddle points. As expected it does not oscillate but generally exhibits a growing behaviour. A transition region around $R_c \sim 1/\sqrt{\Lambda}$ connects both regimes.

\section{Predictions in the Classical Domain}\label{sectcl}

In the previous sections we saw that $\Psi_T$ and $\Psi_{HH}$ oscillate fast in the large overall volume region. This is the realm of superspace where we expect both wave functions to predict a set of real classical histories that are solutions of the Lorentzian Einstein equations \cite{Hartle2008}. In this section we compute this classical ensemble.

Classical evolution emerges at large $R_c$ because both wave functions take a WKB form (more specifically, a sum of such forms) 
\begin{equation}
\Psi[\beta,R_c] \approx  A(\beta,R_c)\exp\{\pm i S(\beta,R_c)/\hbar\} ,
\label{semiclass}
\end{equation}
where $S$ varies rapidly over the region and $A$ varies slowly \cite{Hartle2008}. That is\footnote{We assume for now that $\beta$ is not too small and return to the $\beta \rightarrow 0$ limit in Section \ref{sectbc}.}, each term satisfies
\beq
\label{classcond}
|\nabla _{\beta} A/A| \ll |\nabla _{\beta} S|, \qquad |\nabla _{R_c} A/A| \ll |\nabla _{R_c} S|
\eeq
with $A=\exp{\pm I_R^n}$ and $S=S_{ct}\pm S_R$. 

Under these circumstances the WDW equation implies that $S$ satisfies to a good approximation the classical Hamilton-Jacobi equation  \cite{Hartle2008}. In a suitable coarse-graining the only histories that have a significant probability are then the classical histories corresponding to the integral curves of $S$. This is analogous to the prediction of the classical behavior of a particle in a WKB state in non-relativistic quantum mechanics. 

Integral curves are found by integrating the classical relations relating momenta to derivatives of the action, 
\begin{equation}
\label{momenta}
p_{\beta} = \nabla_{\beta} S, \qquad p_{R_c} = \nabla_{R_c} S.
\end{equation}
The solutions $\beta(t)$ and $R_c(t)$ of \eqref{momenta} are curves in superspace that define a set of classical, Lorentzian histories of the form
\beq \label{lor}
ds^2 = -dt^2 + R^2_c(t) \lp \lp \frac{\beta (t) d\theta}{2\pi}\rp ^2+d\Omega_2^2\rp\,.
\eeq  
The relations between superspace coordinates and momenta \eqref{momenta} mean that to leading order in $\hbar$, and at any one time, the classical histories predicted by a wave function of the universe do not fill classical phase space. Rather, they lie on a surface within classical phase space of half its dimension. 

The relative probabilities of the individual coarse-grained classical histories in the ensemble are given by $A$. They are constant along the integral curves as a consequence of the Wheeler-DeWitt equation (cf \cite{Hartle2008}). Hence they give the tree level measure of different possible universes in the classical ensemble predicted by a particular wave function.

The classical predictions of $\Psi_T$ and $\Psi_{HH}$ evaluated on $S^1 \times S^2$ in the large $R_c$ limit can be obtained from their asymptotic form \eqref{HHresult} and \eqref{Tresult}. It is immediately clear that both wave functions predict the same ensemble of classical histories, albeit with different relative probabilities\footnote{The lack of a divergence as $\beta \rightarrow 0$ in $\Psi_{HH}$ is just one manifestation of this.}. The asymptotic classical ensemble comprises two distinct sets of histories. For $\beta\leq 2\pi / \sqrt{3}$, where $R_0$ is purely imaginary at large $R_c$, the saddle points are associated with classical histories that are simply quotients of de Sitter space. This is because in this region the phase of both wave functions is given entirely by the universal factor $S_{ct}$ given by \eqref{count}. 

By contrast, if $\beta > 2\pi / \sqrt{3}$ then the `renormalised' actions \eqref{lactsad} contribute a phase factor $\pm S_R$ to the wave function
given by \eqref{phase}. In this regime we have 
\beq
\pm \nabla_\beta S_R = -\frac{1}{2} \sqrt{3/\Lambda} \mathrm{Re}[\alpha] \,,
\eeq
and the integral curves specified by 
\bea
\label{intcurves}
\dot{R}_c & = & \frac{1}{\sqrt{\Lambda/3}}\lp \frac{\Lambda}{3}R_c-\frac{1}{2R_c}+\frac{\mathrm{Re}[\alpha]}{2R_c^2}\rp \\
\dot{\beta} & = & \frac{\beta}{\sqrt{\Lambda/3}}\lp \frac{1}{R_c^2} -\frac{3\mathrm{Re}[\alpha]}{2R_c^3}\rp \,.
\eea
Eq. \eqref{intcurves} is nothing but the Lorentzian version of the large $R_c$ expansion of the first integral \eqref{int}. 
The asymptotic solutions are
\beq \label{sol}
\hat{R}(t)\equiv R_c(t) = \exp\left[\sqrt{\frac{\Lambda}{3}}t\right] + {\cal O}(1/\hat{R}^2), \qquad \beta = \beta_{\infty} + {\cal O}(1/\hat{R}^2)
\eeq
where $\beta_{\infty}$ is a constant of integration that specifies the asymptotic relative size of $S^1$ and $S^2$.
Using the first integral \eqref{intcurves} and defining an $S^1$ scale factor $\hat \rho \equiv \dot{\hat{R}}$ the asymptotic Lorentzian solutions \eqref{lor} with $\hat{R}$ and $\beta$ given by \eqref{sol} can be written as 
\beq \label{lsds}
ds^2=-\frac{d\hat{R}^2}{-1+\frac{2M}{\hat{R}}+\frac{\Lambda}{3}\hat{R}^2}+\lp -1+\frac{2M}{\hat{R}}+\frac{\Lambda}{3}\hat{R}^2\rp d x ^2+\hat{R}^2 d\Omega _2^2\,,
\eeq
where the mass $M$ is given by
\beq \label{mass}
M= \frac{1}{2}\mathrm{Re}[\alpha]= \frac{1}{2}\mathrm{Re}[R_0-\frac{\Lambda}{3}R_0^3]\ 
\eeq
and we have introduced hats to distinguish the real variable $\hat R$ that enters in the classical, Lorentzian histories from the complex variable $R$ that describes the saddle point geometries. Therefore at low $\beta$ both wave functions predict (two copies of) an ensemble of Schwarzschild-de Sitter spaces. The black holes have positive mass $M$ in histories associated with saddle points that have $\mathrm{Re}[R_0]>0$. By contrast $\mathrm{Re}[R_0]<0$ saddle points correspond to negative mass black holes. For a quantum gravity theory to be well-defined and stable, configurations like negative mass black holes should be excluded from the physical configuration space \cite{Horowitz1995}. In the context we consider here this means the contributions to the wave function of the corresponding branches of saddle points should be discarded.  This also eliminates the divergence of $\Psi_T$ in the small $\beta$ limit, which (cf. eq. \eqref{divergence}) is associated with the saddle point branch in the second quadrant in Fig. \ref{circle}.

Only part of the Lorentzian geometry is directly (geometrically) connected to the complex saddle points. However one can classically extrapolate the asymptotic solutions using the Lorentzian Einstein equations to obtain other parts of the Lorentzian histories, such as the region inside the horizons where $\hat R$ behaves as a radial direction and $x$ becomes the time direction. To what extent a classical extrapolation beyond the horizon is justified will be discussed elsewhere \cite{Hartle2015}. 

\begin{figure}[t] 
\begin{center}
\includegraphics[scale=0.6]{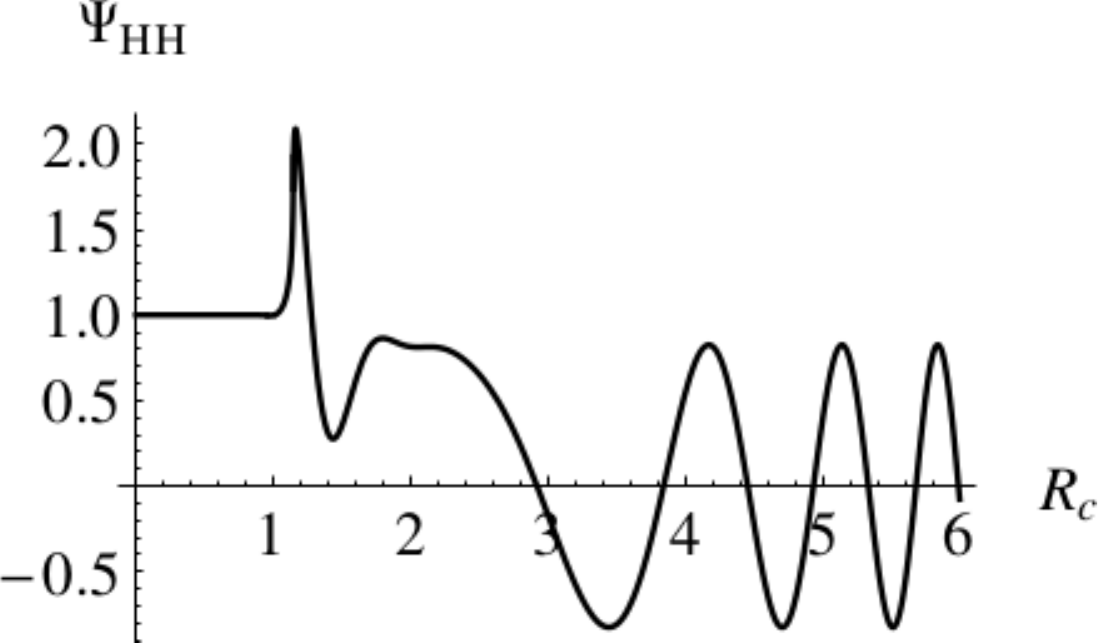}\hspace{1cm}
\end{center}
\caption{Classical evolution only emerges at large boundary radii when $\beta$ is small.}
\label{nocl}
\end{figure}

Finally let us return to the $\beta \rightarrow 0$ regime. In the $\beta \rightarrow 0$ limit the ratio of the gradients \eqref{classcond} with respect to $\beta$ is given by 
\beq
|\nabla _{\beta} I_R|/|\nabla _{\beta} S| \sim  1/\beta^3 R_c^3.
\eeq
This suggests that classical evolution may not be predicted by the wave functions as $\beta \rightarrow 0$, even at large values of $R_c$.
We illustrate this in Figure \ref{nocl}(a) where we show $\Psi_{HH}(R_c)$ for $\beta = 0.1$. One sees that the classical, oscillatory WKB behaviour emerges at boundary radii $R_c$ that are much larger than the radii at which the quantum/classical transition occurs in the wave function evaluated at the larger values of $\beta$ shown in Fig. \ref{nbwfbc}. 

It is interesting that neither $\Psi_T$ nor $\Psi_{HH}$ appears to predict classical evolution in this limit. Moreover this is independent of the inclusion of the divergent branch in $\Psi_T$ in this limit. This is because the breakdown of the classicality conditions as $\beta \rightarrow 0$ is not so much due to large variations in the amplitude but rather to slow variations of the phase. 

The fact that the classicality conditions fail in the $\beta \rightarrow 0$ limit need not itself be an indication of an instability.
However in regions of superspace where the wave function does not predict classical evolution it is difficult to interpret, because probabilities in quantum cosmology are generally assigned to four-dimensional, decoherent classical histories only.

\begin{figure}[t!] 
\begin{center}
\includegraphics[scale=.55]{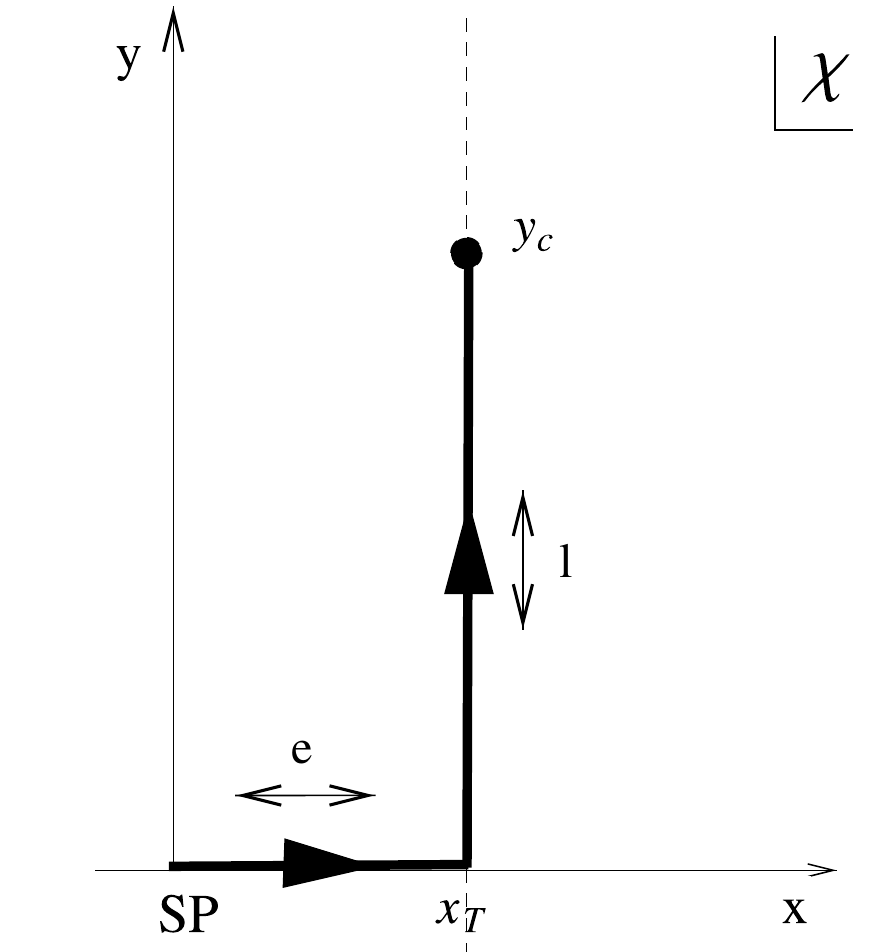}
\end{center}
\caption{Contour in the complex $\chi$-plane along which the geometry of saddle points in the fourth quadrant of Fig. \ref{circle} consists of a complex, approximately Euclidean geometry smoothly joined onto a Lorentzian, classical, Schwarzschild-de Sitter space.}
\label{chiscafact}
\end{figure}

\begin{figure}[t!] 
\begin{center}
\includegraphics[scale=.6]{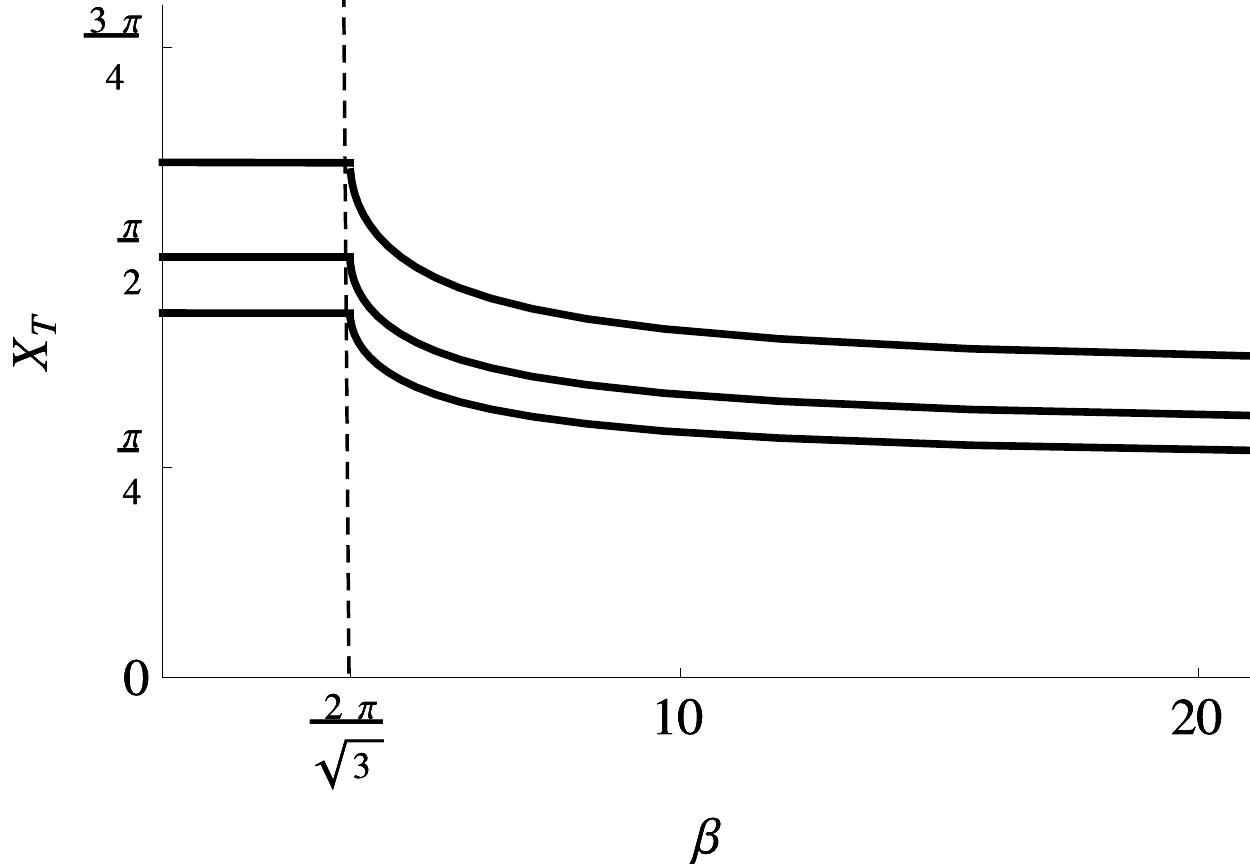} 
\end{center}
\caption{The turning point $x_T$ as a function of $\beta$ for saddle points in the 1st and 4th quadrant of fig. \ref{circle}. From top to bottom $\Lambda = 2, 3, 4$. For $\beta\leq 2\pi/\sqrt{3}$ the turning point is independent of $\beta$ and given by $x_T = \pi/2H$ with $H=\sqrt{\Lambda/3}$ as expected.}
\label{xtvsbeta}
\end{figure}

We have seen that the real Lorentzian histories are not the same as the complex saddle points that determine their probabilities. There is however a convenient geometric representation of the saddle points in which their connection with asymptotically classical histories is explicit \cite{Hartle2008,Hertog2011}. This is obtained by representing the complex solutions $\rho(\chi)$ and $R(\chi)$ along an appropriate contour in the complex $\chi$-plane, starting from $\chi_0$ at the SP to $\chi_c$ at the boundary. An example of such a contour for a saddle point in the fourth quadrant of Fig. \ref{circle} is shown in Fig. \ref{chiscafact}, where we have chosen $\chi_0=0$ without loss of generality. Writing $\chi = x+ iy$ the contour first runs along the real $x$-axis to a turning point $x_T$ from where it goes vertically. The turning point is chosen so that as $y \rightarrow \infty$,
\beq \label{sfds}
\mathrm{Re}[\rho(y)]\rightarrow 0\,, \hspace{1cm} \mathrm{Im}[R(y)]\rightarrow 0 
\eeq
along the contour. A vertical curve of this kind exists in the complex $R_0$-plane provided $R_0$ is a solution of \eqref{quint}. The value of the turning point $x_T$ depends on the boundary data $\beta$ and $\Lambda$. This is illustrated in Figure \ref{xtvsbeta} where we plot $x_T(\beta)$ for three different values of $\Lambda$. 

If the conditions \eqref{sfds} hold then the complex saddle point geometry \eqref{eks} smoothly tends to a real, Lorentzian four-geometry along the vertical part of the contour, of the form
\beq\label{lhis}
ds^2=-d\hat{y}^2+\hat{\rho}^2 d\hat{\omega}^2+\hat{R}^2d\Omega_2^2\,,
\eeq
where $\hat{\rho}(y) \equiv \mathrm{Im}[\rho]$, $\hat{R}(y)\equiv \mathrm{Re}[R]$ and $\hat \omega\equiv \mathrm{Im}[\omega_0]$. This agrees with the classical histories obtained from the integral curves of the phase of the saddle point action. Indeed using the Lorentzian version of the first integral it is straightforward to write the solution \eqref{lhis} in the form \eqref{lsds}.

\section{Holographic Wave Functions}\label{hwfs}

In this section we derive a holographic form of both wave functions on $S^1 \times S^2$ by generalising the results in \cite{Hertog2011} for the Hartle-Hawking wave function on topologically spherical boundaries. 

In \cite{Hertog2011} it was shown that in the large volume limit, the complex saddle points of the Hartle--Hawking wave function on $S^3$ in cosmological models with a positive cosmological constant and a positive scalar potential have a representation in which the geometry consists of a regular Euclidean AdS domain wall that makes a smooth transition to a Lorentzian, inflationary universe that is asymptotically de Sitter. In this representation, the complex transition region between AdS and dS regulates the volume divergences of the AdS action and accounts for the universal phase factor of the wave function. The approximately Euclidean AdS region in turn encodes the information about the state and provides the tree level measure. Specifically the action of all saddle points in this model can be written as
\beq
I_E(\phi, a) = -I^{\rm reg}_{AdS}(\tilde \phi) + iS_{ct}(\phi,a)+{\cal O}(1/a) 
\label{holoHH}
\eeq
where $a$ and $\phi$ are the boundary values of the scale factor and field, and $\tilde \phi$ is a complex intermediate value in the asymptotic AdS region that is fully determined by $\phi$ \cite{Hertog2011}.

Eq. \eqref{holoHH} directly leads to a dual formulation of the {\it semiclassical} Hartle--Hawking measure in which $\Psi_{HH}$ is given in terms of the partition function on $S^3$ of (complex) deformations of the same CFTs that occur in AdS/CFT \cite{Hertog2011}. 

We now show that this result generalises to the Hartle-Hawking wave function evaluated on $S^1 \times S^2$. Further, since the semiclassical tunnelling wave function involves the same saddle points - albeit weighted differently - a similar derivation also yields a holographic form of the tunnelling wave function.

\begin{figure}[t] 
\begin{center}
\includegraphics[scale=.6]{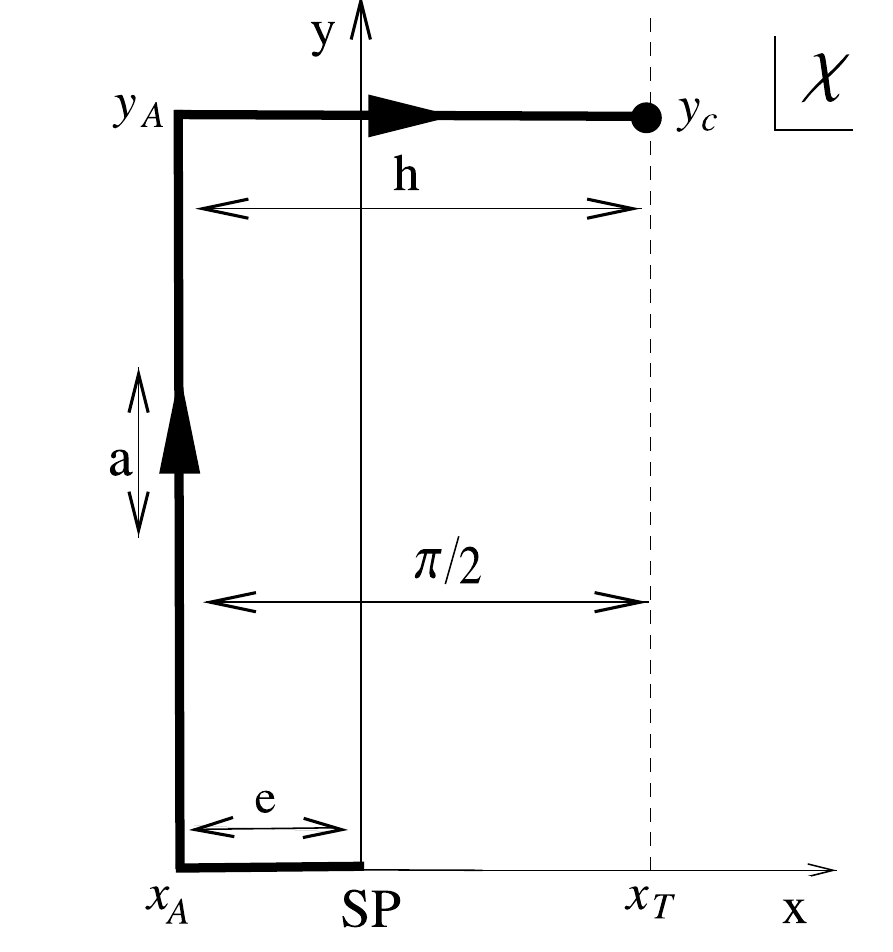}
\end{center}
\caption{The saddle point representation in which the saddle point geometry is Euclidean Schwarzschild-AdS along the vertical part (a) of the contour. This representation serves as a guide in the derivation of a holographic formulation of the Hartle-Hawking wave function in terms of a field theory defined on the conformal boundary geometry.}
\label{holo}
\end{figure}

The action of a saddle point is an integral of its complex geometry that includes an integral over time $\chi$. Different complex contours\footnote{This should not be confused with the choice of contours of path integrals that determines which saddle points to include. We are talking here about different representations of a {\it given} saddle point.} for this time integral yield different geometric representations of the saddle point, without changing the action. This freedom in the choice of contour gives physical meaning to a process of analytic continuation --- not of the Lorentzian histories themselves --- but of the saddle points that through their action define the probability measure on the classical ensemble.   

The contour we considered in Section \ref{sectbc} and shown in Fig. \ref{chiscafact} was particularly useful to exhibit a geometric connection between the complex saddle points and the real, Lorentzian histories. But it is not the only useful representation of the saddle points. Consider the contour shown in Fig. \ref{holo}, for a solution with $\mathrm{Im}[R_0]>0$. This has the same endpoints, the same action, and makes the same predictions. But the geometry is different. The contour can be divided in a part (e) along the horizontal axis from the SP at $\chi_0=0$ to $\chi_e=x_a=x_T - \pi/2$, a part (a) that runs vertically to an intermediate point $\chi_a=x_a+ iy_a$ where $R\equiv =iR_c$ and a part (h) along the horizontal branch connecting $\chi_a$ to the endpoint $\chi_c = x_T+ iy_c$.

The geometry along part (a) is especially interesting. Along this part of the contour 
\beq\label{sfads}
\mathrm{Im}[\rho] \rightarrow 0\,, \qquad \mathrm{Re}[R] \rightarrow 0 \,,\qquad \mathrm{Re}[\omega]\rightarrow 0.
\eeq 
Hence for sufficiently large $y_a$ the saddle point can be conveniently written in terms of approximately real variables,
\beq
\label{ads}
ds^2=-dy^2-\rho^2 d\tilde \omega^2-\tilde R^2 d\Omega_2^2
\eeq
where $\tilde R \equiv -i R $ and $\tilde \omega = -i\omega$. Using the first integral \eqref{int} this becomes 
\beq \label{SAdSrep}
ds^2=-\frac{d\tilde R^2}{1-\frac{2\tilde M}{\tilde R}-\frac{\Lambda_{AdS}}{3}\tilde R^2}-\lp 1-\frac{2 \tilde M}{\tilde R}-\frac{\Lambda_{AdS}}{3}\tilde R^2\rp d\tilde \omega ^2-\tilde R^2d\Omega _2^2\,.
\eeq
This is Euclidean Schwarzschild -- AdS with $\Lambda_{AdS} = -\Lambda$ and mass
\beq
\label{massads}
\tilde M \equiv -\frac{i}{2}\alpha = -\frac{i}{2}(R_0+\frac{\Lambda_{AdS}}{3}R_0^3) = \frac{1}{2}(\tilde R_0 - \frac{\Lambda_{AdS}}{3}\tilde R_0^3)
\eeq
with $R_0$ given by \eqref{saddlesl}. 

Fig. \ref{circle} shows that the AdS black hole mass is real and positive in the large volume limit for $\beta \leq 2\pi/\sqrt{3}$ whereas it is complex for larger $\beta$. This is expected since there is a critical temperature $T_c = \sqrt{\Lambda}/(2\pi)$ below which there are no (real!) AdS black holes \cite{Hawking1982}. Evaluating eq. \eqref{ombound} in the large $R_c$ limit shows that $\beta=2\pi/\sqrt{3}$ precisely corresponds to $T_c$. The saddle point branch outside the $|R_0| = 1/\sqrt{\Lambda}$ circle in Fig. \ref{circle} corresponds to the large black holes that are  thermodynamically stable in AdS. 

 A similar AdS representation can be found for saddle points with $\mathrm{Im}[R_0] <0$. In this case the vertical part (h) of the contour runs down to negative values of $y$. The Schwarzschild-AdS geometry \eqref{SAdSrep} of these saddle points can be made explicit in terms of the radial variable $\tilde R \equiv iR$ which is real and positive along the AdS part of the contour. The black hole mass $\tilde M =+\frac{i}{2}\alpha$ which is again given by the right hand side of \eqref{massads} and positive.

It remains to compute the action along the AdS contour of Fig. \ref{holo}, in the large volume limit $R_c \gg 1/\sqrt{\Lambda}$.
The total action integrated along the first two legs of the contour is given by
\beq \label{holoact}
I^{(e)+(a)}_E(\beta, \tilde R_c) =- \frac{\beta}{\sqrt{\Lambda/3}}\lp \frac{\Lambda_{AdS}}{3} \tilde R_c^3 -\frac{1}{2} \tilde R_c
-\frac{i}{4} R_0(1-\frac{\Lambda_{AdS}}{3}R_0^2) \rp +\mathcal{O}\lp \frac{1}{\tilde R_c}\rp \,.
\eeq
As expected the action along (a) exhibits the usual volume divergences in the $\tilde R_c \rightarrow \infty$ limit that are characteristic of the action of asymptotically AdS spaces. The divergent terms are universal and account for what are known as the counterterms \eqref{count} in holographic discussions \cite{Emparan1999,Skenderis2002}. The asymptotically finite contribution to the action along (a) is not universal and encodes information about the state and the dynamics. It is closely related to the regularised AdS action which in terms of the natural radial AdS variable $\tilde R$ reads
\beq
I_{AdS}^{\rm reg}(\beta)= \frac{\beta}{4\sqrt{\Lambda/3}} \tilde R_0(1+\frac{\Lambda_{AdS}}{3}\tilde R_0^2)\ .
\eeq
Substituting this in \eqref{holoact} yields
\beq \label{holoact2}
I^{(e)+(a)}_E(\beta, \tilde R_c) =- I_{AdS}^{\rm reg}(\beta)  +S_{ct}(\beta,\tilde R_c) +\mathcal{O}\lp 1/\tilde R_c\rp \,, 
\eeq
where the counterterms $S_{ct}$ are given by \eqref{count} evaluated on the boundary $(\beta,\tilde R_c)$.
The action along the horizontal leg of the contour is given by
\beq \label{holoact3}
I^{(h)}_E(\beta, \tilde R_c,R_c)= -S_{ct}(\beta,\tilde R_c)-iS_{ct}(\beta, R_c)\,.
\eeq
Hence this part of the contour merely regulates the volume divergences and supplies the universal phase factor of the wave function. 
Taken together this means the saddle point actions in the large volume limit can be written as
\beq
\label{holoact4}
I_E (\beta,R_c) \approx - I_{AdS}^{\rm reg}(\beta) -iS_{ct}(\beta, R_c)\,.
\eeq

Therefore the requirement that a configuration on the final boundary behaves classically, with constant $I_R$, automatically regulates the volume divergences associated with the action of the Euclidean AdS regime of the saddle point. This implies that the leading order in $\hbar$ probabilities of the classical Schwarzschild-de Sitter histories can be calculated either from the dS representation of the saddle points or from their representation as Euclidean Schwarzschild-AdS spaces.

Eq. \eqref{holoact4} provides a natural connection between $\Psi_{HH}$ on $S^1 \times S^2$ boundaries in pure de Sitter gravity and Euclidean AdS/CFT. The Euclidean AdS/CFT correspondence relates $I^{\rm  reg}_{AdS}(\beta)$ in turn to minus the logarithm of the large $N$ limit of the partition function $Z_{CFT}[\beta]$ of a dual conformal field theory defined on the conformal boundary $\gamma$ (cf. eq. \eqref{boundl}). This yields a dual formulation of the semiclassical NBWF -- and hence a concrete realization of a semiclassical dS/CFT duality -- in terms of one of the known, unitary dual field theories familiar from AdS/CFT \cite{Hertog2011}. In this dual description, the semiclassical Hartle-Hawking wave function \eqref{HHresult} is of the form
\beq \label{dualHH}
\frac{1}{Z_{CFT}(\beta)} \cos [S_{ct}(\beta,R_c)]\,.
\eeq
Hence the argument $\beta$ of the wave function in the large volume limit enters as an external source in the dual partition function. The dependence of the partition function on the value of $\beta$ then gives a dual Hartle-Hawking probability measure on the space of classical asymptotic configurations.

The semiclassical tunnelling wave function \eqref{Tresult} involves the same complex saddle points as the HH wave function but weighted differently. Therefore \eqref{holoact4} can also be used to put forward a holographic form of $\Psi_T$ in the large $R_c$-regime \cite{Conti2015}. Substituting \eqref{holoact4} in \eqref{Tresult} yields the following holographic representation of the growing branch of $\Psi_T$,
\beq \label{dualT}
\Psi_T \sim Z_{CFT}(\beta) e^{iS_{ct}(\beta,R_c)}.
\eeq

In principle one can use the holographic expressions \eqref{dualHH} and \eqref{dualT} to compare semiclassical bulk results for the wave functions with the predictions of a dual boundary theory. At this point explicit boundary calculations with scalar sources are feasible only for the $Sp(N)$ or $O(N)$ conformal field theories which are conjectured to be dual to Vasiliev's higher-spin gravity, respectively in asymptotic de Sitter space \cite{Anninos2011} and in AdS \cite{Klebanov2002}.

Nevertheless the partition functions of those models might qualitatively capture certain aspects of the behaviour of wave functions in Einstein gravity. This was the approach pursued in \cite{Anninos2012,Banerjee2013} in the context of the dS/CFT proposal of \cite{Anninos2011}, where it was found that the partition function of the $Sp(N)$ model on $S^1 \times S^2$ exhibits a divergence in the small $S^1$ limit that is of the same form as the behaviour of $\Psi_T$ in Einstein gravity, {\it provided the second branch is included}\footnote{In \cite{Anninos2012,Banerjee2013} the divergence in the $Sp(N)$ model was associated with a divergence of the Hartle-Hawking wave function in Einstein gravity. However Section 2 of this paper illustrates that the calculations in \cite{Anninos2012,Banerjee2013} actually compute the tunnelling wave function.}.

The holographic form of the wave functions we derived above is differs somewhat from the dS/CFT proposal \cite{Anninos2011} used in \cite{Anninos2012,Banerjee2013} in that it (again for Vasiliev gravity) involves the AdS dual $O(N)$ model rather than the $Sp(N)$ model. This is because we have used the complex analytic structure of the saddle points to relate the semiclassical wave functions in asymptotic dS to (Euclidean) AdS rather than taking $N \rightarrow -N$ in the dual to `continue' from AdS to dS. However the net result at this level of comparison is the same. Indeed the partition function of the bosonic $O(N)$ vector model at large temperature is \cite{Shenker2011}
\beq \label{Z}
\log Z_{CFT} = 4\zeta(3) N/\beta^2\,.
\eeq
Substituting this in \eqref{dualT} one sees this qualitatively reproduces the behaviour \eqref{divergence} of $\Psi_T$ along the divergent branch in the small $S^1$ limit.

It follows from \eqref{dualHH} that the dual partition function \eqref{Z} also qualitatively reproduces the behaviour of $\Psi_{HH}$ along the same branch. In the Hartle-Hawking state however this is a subleading branch of the wave function. It is an important open problem in holographic cosmology and in holography in general to access the second branch of bulk wave functions from a dual boundary theory.

\section{Discussion}\label{disc}

We have evaluated the semiclassical tunneling and Hartle-Hawking wave functions on $S^1 \times S^2$ in Einstein gravity coupled to a positive cosmological constant. Over most of superspace there are four branches of complex saddle points that can contribute to the wave functions. In the classical region of superspace -- at large overall volume -- the wave functions predict an ensemble of real Lorentzian histories. Asymptotically the wave functions are functions of the relative size of $S^1$ and $S^2$. When the $S^1$ is sufficiently large (relative to the $S^2$) the real, asymptotic classical histories that correspond to the complex saddle points are Schwarzschild-de Sitter black holes. Two branches describe black holes with positive mass whereas the remaining two are associated with negative mass black holes. The latter branches give rise to a divergence in the tunneling state at small $S^1$. By contrast the Hartle-Hawking wave function appears to be well-behaved in this regime.

Singularities associated with negative mass black holes are not expected to be `resolved' in quantum gravity. This is because the resolution of such singularities would yield regular solutions with negative energy and thus lead to a theory without a stable ground state \cite{Horowitz1995}. Configurations like negative mass black holes with naked, timelike singularities should therefore be excluded from the physical configuration space. This is possible if they lie in a separate `superselection' sector of the theory. In the quantum cosmological context considered here the most natural way to do this is to discard the contribution 
to the wave function of the branches of saddle points associated with negative mass black holes at large $S^1$. We have shown this also resolves the problem of the divergence of the tunneling state in Einstein gravity in the small $S^1$ limit.

Whether this is the correct procedure in Vasiliev gravity remains an open question, because Vasiliev gravity may not be a stable theory. At present most of what we know about Vasiliev gravity is based on calculations in the boundary theory. One might argue that our results in the context of an Einstein gravity bulk support the interpretation \cite{Anninos2012,Banerjee2013} that the divergences indicate Vasiliev gravity is unstable, because we have shown they are associated with negative mass black holes in de Sitter space.

\vskip 1cm
\centerline{\bf Acknowledgements}
\vskip 1cm
We thank Dionysios Anninos, Frederik Denef, James Hartle, Gary Horowitz, Liu Lihui, Ellen van der Woerd and Alex Vilenkin for useful discussions. This work is supported in part by the National Science Foundation of Belgium (FWO) grant G.001.12 Odysseus. TH is supported by the European Research Council grant no. ERC-2013-CoG 616732 HoloQosmos. We also acknowledge support from the Belgian Federal Science Policy Office through the Inter-University Attraction Pole P7/37 and from the European Science Foundation through the `Holograv' Network.


\end{document}